\documentclass[aps,prl,preprint,superscriptaddress]{revtex4}
\usepackage{amsmath}
\usepackage{amssymb}
\usepackage{epsfig}
\usepackage{color}
\usepackage{graphicx}
\usepackage{graphicx,amsmath}
\usepackage{bm}
\newcommand{\beq}{\begin{equation}}
\newcommand{\eeq}{\end{equation}}
\begin{document}
\title{Experiments in machine learning of $\alpha$-decay half-lives}
\author{Paulo S.~A.~Freitas}
\affiliation{FCEE, Departamento de Matem\'{a}tica,
        Universidade da Madeira, 9020-105 Funchal, Madeira, Portugal}
\author{John W.~Clark}
\affiliation{McDonnell Center for the Space Sciences \&
	Department of Physics, Washington University,
	St.~Louis, MO 63130, USA}
\affiliation{Centro de Investiga\c{c}\~{a}o em Matem\'{a}tica
	e Aplica\c{c}\~{o}es, University of Madeira, 9020-105
	Funchal, Madeira, Portugal}
	
\begin{abstract}
Artificial neural networks are trained by a standard backpropagation 
learning algorithm with regularization to model and predict the
systematics of -decay of heavy and superheavy nuclei.  
This approach to regression is implemented in two alternative modes: 
(i) construction of a statistical global model based solely on
available experimental data for alpha-decay half-lives, and (ii)
modeling of the {\it residuals} between the predictions of
state-of-the-art phenomenological model (specifically,
the effective liquid-drop model (ELDM)) and experiment.  Analysis 
of the results provide insights on the strengths and limitations of 
this application of machine learning (ML) to exploration of the 
nuclear landscape in regions beyond the valley of stability.
\end{abstract}


\maketitle

\section{1. Introduction}

Machine learning has been applied to global statistical modeling
and prediction of nuclear properties since the early 1990's 
\cite{ainp}. Relative to phenomenology and fundamental theory, these 
studies have yielded complementary or competitive results for atomic 
masses, nucleon separation energies, charge radii, ground-state spins and 
parities, branching ratios, and beta-decay half-lives.  Recently, this 
original wave has been followed up by a major surge of activity swept along 
by current enthusiasm for machine learning techniques across many scientific 
disciplines, notably in condensed matter physics and quantum chemistry 
(see especially Refs.~\cite{utama,zhang,negoita,neufcourt,niu}, and
work cited therein).  Although there have been significant improvements 
in machine-learning techniques, the driving force has been the vast 
increase in computational power during the last two decades.

Modeling and prediction of $\alpha$-decay observables of heavy nuclei has 
not received much attention within this thrust, past or present, although
the study of superheavy nuclei, having $Z > 104$, remains an important
dimension of experimental and theoretical study in expanding our knowledge 
of the nuclear landscape well beyond the familiar valley of stability 
\cite{organ}. Significantly, $\alpha$ decay and fission are the dominant
modes of decay for this unique class of nuclei.  Indeed, superheavy 
nuclei so defined are of special interest in that their existence 
is contingent on shell effects, giving rise to the ``magic island'' 
of nuclei enjoying relative stability.  At this point, synthesis of many 
superheavy elements and isotopes has been achieved through cold, warm, 
and hot fusion, up to $Z = 118$.

Advances in fundamental understanding of the $\alpha$-decay process
date back to the fundamental breakthrough of Gamow and Condon and 
Gurney, who identified the underlying mechanism as quantum tunneling.  
Semi-empirical formulas have been introduced to estimate $\alpha$-decay 
half-lives, such as that of Viola and Seaborg \cite{viola} and much 
more recently those of Royer \cite{royerform} and Denisov \cite{denisov}.  
However, apart from efforts toward true microscopic description of 
$\alpha$ decay, the current state-of-the art for its quantitative 
theoretical treatment resides in two phenomenological models based 
on semi-classical considerations.  These are (i) the generalized 
liquid-drop model (GLDM) \cite{remaud,cui16}, in fission-like and 
cluster-like modes, and (ii) the effective liquid-drop model (ELDM) 
\cite{goncalves,duarte}, together with their refinements.  Among 
recent developments, see especially Refs.~\cite{cui18,cui19,santhosh}.

Past and recent efforts toward machine learning of $\alpha$ decay
have been limited to Refs.~\cite{bayram,rodriguez}.  The first of 
these works was an exploratory study indicating that an artificial
neural network approach to this problem domain may be successful.  The
second (which contains a very useful update of current ML activity
in nuclear physics) concentrates on machine learning of the energy 
release $Q_\alpha$, an essential input to calculation of $\alpha$-decay 
half-lives.

Our aim in the present work is two-fold, namely to develop stand-alone 
statistical models for the $\alpha$-decay process, with input limited 
principally to $N$, $Z$, and $Q_\alpha$, and to refine the ELDM model 
by developing neural-network models of the residuals between this 
model and experiment.  We begin with a brief introduction to machine 
learning, as realized by feedforward multilayer neural networks.

\section{2. Elements of Statistical Modeling}

Our objective is to use machine learning (ML) to model the mapping 
from the nucleonic content $(N,Z)$ of a given nuclear species to
the $\alpha$-decay half-life $T_{1/2}$ of that nuclide, based on 
a set of experimental data that provides examples of this 
mapping.  Since the goal is a real number rather than a correct
choice among discrete outcomes, one is dealing with a task of 
regression (as opposed to classification), such that the problem
may be formulated as follows.

The mathematical problem of approximating a function $f$ 
that maps an input variable $x$ to an output variable $y$ may be 
expressed as
\beq
y  = f(x,\theta;\Upsilon) + \epsilon,
\label{regression}
\eeq
with a set of data points $(y_n,x_n)$, $n=1,\ldots,n_T$ (i.e., $n_T$ 
given training ``patterns'') being made available for interpolation and
extrapolation. As indicated, the function $f$ depends on unknown 
parameters $\theta$ that are to be estimated.  In general, the 
mapping also depends on certain hyperparameters denoted $\Upsilon$, 
which must be carefully specified {\it a priori}, before estimation 
of $\theta$.  The term $\epsilon$ represents noise, a residual
stochastic component of the model.

In applications of machine learning to nuclear physics, the primary goal
is reliable extrapolation from the existing database of a given nuclear 
property, i.e., {\it generalization}, rather than a precise representation 
of the training data itself.  With generalization the dominant
consideration, the complexity of the model must be controlled as 
well as optimized, seeking the best compromise between simplicity and 
flexibility.  This is the model selection problem.  If the model is 
too simple or inflexible, it may not fit the data well enough, resulting 
in poor predictive power. Conversely, if it is too flexible relative 
to the available training data, it is susceptible to overfitting of 
that data (which may be contaminated with spurious noise), at the 
expense of extrapability to new data regimes. 

In other words, apart from choosing an appropriate machine-learning
framework -- for example, multilayer feedforward neural networks, 
radial basis function representations, support vector machines, Bayesian 
neural nets -- the ability to find the right hyperparameters for the assumed 
statistical model can be crucial for its performance based on the existing 
data.  Grid searches can be employed in seeking the most favorable 
combinations of hyperparameter values.

A valuable approach to controlling the complexity of a model, i.e.,
avoiding over-parameterization, is {\it regularization}.  This
involves adding a penalty function to the assumed loss (or ``cost'') 
function (which is analogous to energy, being subject to 
minimization in optimization of the model). Scaling such a penalty 
term with an adjustable multiplicative parameter allows for control 
of the degree of regularization. In seeking a model having strong 
generalization capability, it must be expected that such
a model will typically {\it not} give the smallest error with
respect to the given training data.  There is a natural trade-off
between training error and generalization error.

The most straightforward approach to model selection is to evaluate
the performance of model candidates on data {\it independent} of that
used in training, which serves as a {\it validation set}.  Using the 
original training set, one may generate as many such candidates as there 
are different combinations of model hyperparameters $\Upsilon$ (or a chosen 
subset of them).  The models of this collection are then evaluated 
based on their performance on the validation set, the model with the best 
score being selected.  Since there will still be some dependence between 
the chosen model and the assumed validation set, overfitting to this set 
may occur.  Therefore the performance of the selected model should be 
confirmed by evaluating it on a third data set that is independent of both 
training and validation sets: the {\it test set}.

The validation process can be resource intensive.  In practice,
the quantity of data available may not permit the luxury of reserving
a significant portion for model selection.  In such cases, 
$\cal X$-{\it fold cross-validation} provides an alternative. The training
data is partitioned randomly into $\cal X$ approximately equal subsets.
Considering in turn each of these subsets as the validation (or
``holdout'') set and the remaining ${\cal X} -1$ subsets as comprising
the training set, a validation score is assigned to the resulting
model.  The completed process yields $\cal X$ competing models, and
the final cross-validation error is determined from the mean
validation score over the $\cal X$ decompositions.  Typical choices
of $\cal X$ are 5 to 10.  Before computing the generalization error,
the selected model is retrained using the {\it full} training set.

\section{3. Multilayer Perceptrons (MLP) and their Training}

The artificial neural network learning algorithm adopted for our study 
of $\alpha$ decay of heavy nuclei is based on the feedforward 
multilayer perceptron architecture sketched in Fig.~1, consisting of an 
input layer, an output layer, and one or more intermediate (``hidden'') 
layers, each composed of ``nodes'' or neuron-like units.
Information flows exclusively from input to hidden to output
layers.  For simplicity, we consider here only one hidden layer, 
consisting a number $J$ of nodes $j$ that serves as a hyperparameter. 
(Generalization to multiple hidden layers is straightforward).  
Ordinarily, every node in a given layer provides input to every 
node in the succeeding layer, with no reciprocal connections;
in this respect the network is fully connected.

The input layer is composed of $I$ nodes $i$ that serve only as 
registers of the activities $x_i$ of each input pattern.  However,
the $J$ nodes $j$ of the hidden layer and the $K$ nodes $k$ of the
output layer process, in turn, their respective inputs $x_i$ and 
$h_j$ in a manner typically analogous to biological neurons. This 
produces corresponding outputs $h_j = f(b_j + \sum_i w_{ji}x_i)$ and 
\beq
y_k = b_k + \sum_j w_{kj}h_j  
\eeq
in terms of $IJ + JK$ connection weights plus $J+K$ bias parameters: 
collectively $w = \{w_{ji},w_{kj}\}$ and $b = \{b_j,b_k\}$.  For the 
application reported here, the activation (or ``firing'') function $f$ 
of processing units in the middle layer was chosen as $f(u) = \tanh (u)$, 
resembling the response of biological neurons.  

Here we must note that because regression is the main task to be performed in 
our application, the outputs are not required to lie in any particular 
interval.  Accordingly, outputs of the network are better computed as 
linear combinations of the outputs of the hidden layer plus a bias. 
This is the same as not considering any activation function in the
output layer.

The collection $w$ of weights of the connections from input to hidden 
units and from the hidden units to the output units, together with the 
biases $b$ of these units, determine the given MLP as a proposed solution 
of the regression problem represented by Eq.~(\ref{regression}).  
In the {\it backpropagation learning algorithm}, the weights are commonly 
optimized by gradient-descent minimization of the loss function
\beq
L(w,\lambda) = \frac{1}{2} \sum_{n=1}^{n_T}[y_n - f(x_n,w)]^2 
+ \frac{1}{2}\lambda ||w||^2,
\eeq
which is constructed from a sum of squares of output errors for the
$n_T$ examples provided by the training set, plus a simple 
regularization term that punishes weights of large magnitude. In 
our implementations of gradient descent, large weight oscillations 
are discouraged by a ``classical'' momentum term
\cite{qiang}
in each training step $s$, according to 
\beq
\Delta w_s = -\alpha \nabla L(w_{s-1}) + \mu \Delta w_{s-1} 
\eeq
(or alternatively with Nesterov's momentum \cite{nesterov}). 
Here $\nabla L_w$ denotes the gradient of the loss function 
evaluated at $w_s$, while $0 < \alpha < 1$ is the learning rate and 
$0 \leq \mu < 1$ is the momentum parameter.  (Also implemented in 
our applications are adaptive moments (``Adam'') \cite{kingma}, involving 
moving averages of gradient and squared gradient, bias correction 
for second moments, and associated weight updating.)

We refer the reader to the standard texts \cite{haykin,bishop,neal}
for more thorough and authoritative introductions to neural networks
and machine learning.

\section{4. Specialization to Machine Learning (ML) of Alpha Decay}

In our exploratory study, we focus attention on one specific application 
of ML to statistical modeling and prediction of nuclear properties, namely 
$\alpha$-decay lifetimes of heavy nuclei, based on the multilayer perceptron 
procedures outlined in the preceding section.  Additionally, some results 
of alternative strategies for exploration of the same problem domain
will be summarized.  

For this ML application, the data involved in both ML statistical modeling
and assessment of the theoretical model chosen as reference include:
\begin{itemize}
\item[(i)]
The experimental half-lives $T^{\rm exp}_{1/2}$ (measured in seconds) 
of a set of 150 nuclear isotopes $(N,Z)$ with $Z \geq 104$. 
\item[(ii)] 
Corresponding experimental values of the energy release $Q_\alpha(Z,N)$,
which is $M(Z,N) -M(Z-2,N-2) - M(2,2)$ in terms of reactant masses. 
\end{itemize}
These data sets, assembled from Refs.~\cite{cui18,cui19}, contain entries 
for the 120 neutron-deficient nuclides in Table I of Ref.~\cite{cui19}, 
supplemented by entries for the 30 non-overlapping nuclides with 
$Z=80-118$ drawn Table I of Ref.~\cite{cui18}.  (Sources for the 
experimental values are cited in these two studies; such values and 
their uncertainties may differ from those provided by the National 
Nuclear Data Center.)  Fig.~2 displays plots of the experimentally 
determined $\log T_{1/2}$ values versus $Z$, $N$, $A$, and input 
$Q_\alpha$, for the composite data set of 150 nuclides.

Inputs to the MLP network models include $Z$, $N$, their parities,
and their distances $d_Z$ and $d_N$ from respective proton
and neutron magic numbers, viz.\ 2, 8, 20, 28, 50, 82, and 126
for $Z$, plus 184 for $N$, for a total of 7 inputs, including the
experimental $Q_\alpha$ values.

As is conventional in modeling of decay rates, the actual quantity 
targeted is the base-10 logarithm $t = \log_{10} T_{1/2}$ of the
half-life $T_{1/2}$.  For a given model of the half-life data (whether 
theoretical or statistical), the key figure of merit is the smallness of 
the standard deviation
\beq
\sigma = \frac{1}{n_T}\left[\sum_{n=1}^{n_T} (t_n^{\rm exp} - t_n^{\rm mod})^2
\right]^{1/2}
\label{sigdef}
\eeq
of the model estimates from experiment.  If the ML procedure is 
used in the alternative mode of modeling the {\it differences} 
between a given theoretical model (th) and experiment (exp), 
so as to develop a statistical model of the ``residuals'' (res), the 
second term within the parentheses in Eq.~\ref{sigdef} is 
replaced by $-t_n^{\rm th} - t_n^{\rm res}$.

Inputs to the neural network model, $I=7$ in total, consist of the
atomic and neutron numbers $Z$ and $N$, their parities, the decay 
energy $Q_\alpha$, and the distances $|Z-Z_m|$ and $|N-N_m|$ 
of a given input nuclide from the nearest respective magic numbers 
(viz. 8, 20, 28, 50, 82, 126 for protons, plus 84 for neutrons).

For model assessment, the full data set was randomly split into: 
\begin{itemize}
\item[(i)]
A {\it training set} (TR), comprised of 80\% of the data 
(120 nuclear ``patterns''.
\item[(ii)]
A {\it test set} (TS), 20\% of the data (30 patterns), provided by 
the remainder.
\end{itemize}
Model selection was carried out by five-fold cross-validation, as 
explained in the preceding section.  A total of 2750 combinations of 
$(J, \lambda, \alpha, \mu)$ values (i.e.\ number of hidden units, 
regularization strength, learning rate, and momentum) were 
compared.  The specific learning algorithm applied was gradient 
descent with Nesterov's momentum.  The number of epochs (passages 
through the training data) was 3000.  Batch updating 
(batch size = 32) was applied, with data shuffling.  During the learning phase, 
the quality measure $\sigma$ was monitored at each epoch. Weights/biases 
were chosen to minimize $\sigma(TR)$ over epochs.  These network 
parameters were then used to make predictions for new nuclear examples, 
outside the training set.

Two kinds of network model were constructed:
\begin{itemize}
\item[(i)]
{\bf Net1}: Trained on the experimental half-life data for the selected 
150 nuclear examples drawn from Refs.~\cite{cui19,cui18}, yielding
a purely statistical model of $\alpha$ decay.  Fig.~3 provides different
visualizations of the training and test sets used for this network
model.
\item[(ii)]
{\bf Net2}: Trained on the data set consisting of the {\it differences}
between the predictions of a given theoretical model of $\alpha$ decay
(specifically, ELDM), providing statistically derived corrections 
to this model.
\end{itemize}
After finding the best combination $(J^*,\lambda^*,\alpha^*,\mu^*)$ of 
hyperparameters for {\bf Net1} and {\bf Net2}, the most favored nets 
were trained again, but now using the whole set of training data at once.  

\section{5. Results for Learning and Prediction}

Table 1 summarizes the results of these machine-learning procedures
applied to $\alpha$-decay half-life data for the 150 heavy nuclei extracted
from Refs.~\cite{cui19,cui18}.  Optimal sets of the parameters
$(J,\lambda,\alpha,\mu)$ are shown for the two network models,
along with the respective quality measures $\sigma$ attained by these 
nets on training, validation, and test sets, respectively denoted
TR, VS, and TS.  Corresponding values of the standard-deviation quality 
measure $\sigma$ as achieved with the theoretical models employed 
by Cui et al.\ in these references are entered for comparison.   
Fig.~4 gives a schematic view of the quality of prediction achieved by 
{\bf Net1} and {\bf Net2}.  Fig.~5 provides more refined graphical 
representations of the performance of {\bf Net2} on training and test 
sets.

{\bf Table 1}.  Summary of machine-learning results for the quality measure 
$\sigma$, compared with corresponding results from the theoretical 
models applied in Refs.~\cite{cui19,cui18}. Also displayed are the 
optimal parameter choices used for the ML models.
\begin{center} \begin{tabular}{||c c c c c ||} 
\hline 
{\rm Predictor}  & $\sigma({\rm TR})$ ~ & $\sigma({\rm VS})$  
~ & $\sigma({\rm TS})$ &  $(J^*,\lambda^*,\alpha^*,\mu^*)$\\ [0.5ex] 
\hline
 {\rm ELDM}& 0.5690 ~ & -- ~ & 0.5845 ~ & -- \\ 
 {\bf Net1}  & 0.4468 ~ & 0.5073 ~ & 0.4910 ~ & $(4,10^{-1},10^{-3},0.99)$  \\
 {\bf Net2}  & 0.4586 ~ & 0.4756 ~ &  0.4773 ~& $(20,1,10^{-3},0.99)$ \\
\hline
\end{tabular}
\end{center}

\section{Conclusions and Outlook}

Table 1 shows a modest reduction of the error measure $\sigma$
of the theoretical models upon implementing machine learning, 
although it is clear that ML is quantitatively competitive with 
the best of current quantum-mechanical models of $\alpha$ decay.  
It is significant, however, that quite unlike what is found in 
ML of nuclear masses \cite{ainp,utama} there is no appreciable 
improvement relative to the performance of the purely statistical 
ML model provided by ${\bf Net1}$, when machine learning is applied 
to create ${\bf Net2}$ by training only on the residuals of the 
chosen theoretical model.

There are several possible explanations for these findings. A 
major consideration is the paucity of data compared to the machine 
learning of nuclear masses.  For $\alpha$ decay, the examples available
for training are at most of order hundreds, whereas for masses
there between two and three thousand measurements that may be
used for training.  It was for this reason that we chose to enlarge 
the $\alpha$-decay data set of Ref.~\cite{cui19} by adding non-overlapping 
examples from the data set of Ref.~\cite{cui18}. This may have been 
counterproductive in that all of the added nuclei have $Z > 104$, 
whereas the data set of Ref.~\cite{cui19} is made up of neutron-deficient 
nuclei.  More broadly, the irreducible sparsity of training data 
increases the danger of overlearning and accentuates the sacrifices 
made in its avoidance.   

These considerations would not be so serious were it not for the fact
that the experimental measurements of $\alpha$ and $\beta$ decay rates are
subject to relatively large errors compared to masses.  This is 
responsible for the accepted rule-of-thumb that $\alpha$-decay 
half-life predictions (measured in seconds) within an order of 
magnitude of the true value are considered acceptable (cf.\ the 
"hindrance factor").  The clear implication is that in ML applications 
to nuclear decay, the learning algorithm should give explicit 
consideration to the noisy component of the data, to the extent it 
is quantifiable.  A straightforward modification of the backpropagation 
algorithm to cope with this situation has been developed and tested in 
Ref.~\cite{cleanprop}, and will be implemented in further ML 
treatments of $\alpha$ decay.

The otherwise unexpected results of an additional experiment carried
out as a limiting case are relevant to the situation described.  We sought
to answer the question:  How well does a degenerate version of the
network model considered here, i.e., the Elementary Perceptron, 
perform on the alpha-decay problem?  The answer is contained in
Table 2, where {\bf Net1} and {\bf Net2} now have only a single 
processing layer (the output layer).

{\bf Table 2}. Performance of the Elementary Perceptron 
in learning and prediction of alpha-decay half-lives (cf. Table 1).
Also displayed are the optimal parameter choices used for the ML models.
\begin{center} \begin{tabular}{||c c c c c ||} 
\hline 
{\rm Predictor}  & $\sigma({\rm TR})$ ~ & $\sigma({\rm VS})$  
~ & $\sigma({\rm TS})$ &  $(J^*,\lambda^*,\alpha^*,\mu^*)$\\ [0.5ex] 
\hline
 {\bf Net1}  & 0.4840 ~ & 0.5309 ~ & 0.5153~ & $(0,0,10^{-2},0.0)$  \\
 {\bf Net2}  & 0.4399 ~ & 0.4837 ~ &  0.4638 ~& $(0,0,10^{-2},0.99)$ \\
\hline
\end{tabular}
\end{center}

These results are not inconsistent with the above discussion.  In 
the context of the alpha-decay machine-learning problem where 
uncertainties in measured half-lives are large, course data
can be modeled just as well by a course model.

\section{Acknowledgments}

JWC expresses his gratitude for the hospitality of the University
of Madeira during extended residence, and continuing support 
from the McDonnell Center for the Space Sciences.

\newpage


\begin{figure}
\includegraphics[width=0.6\textwidth]{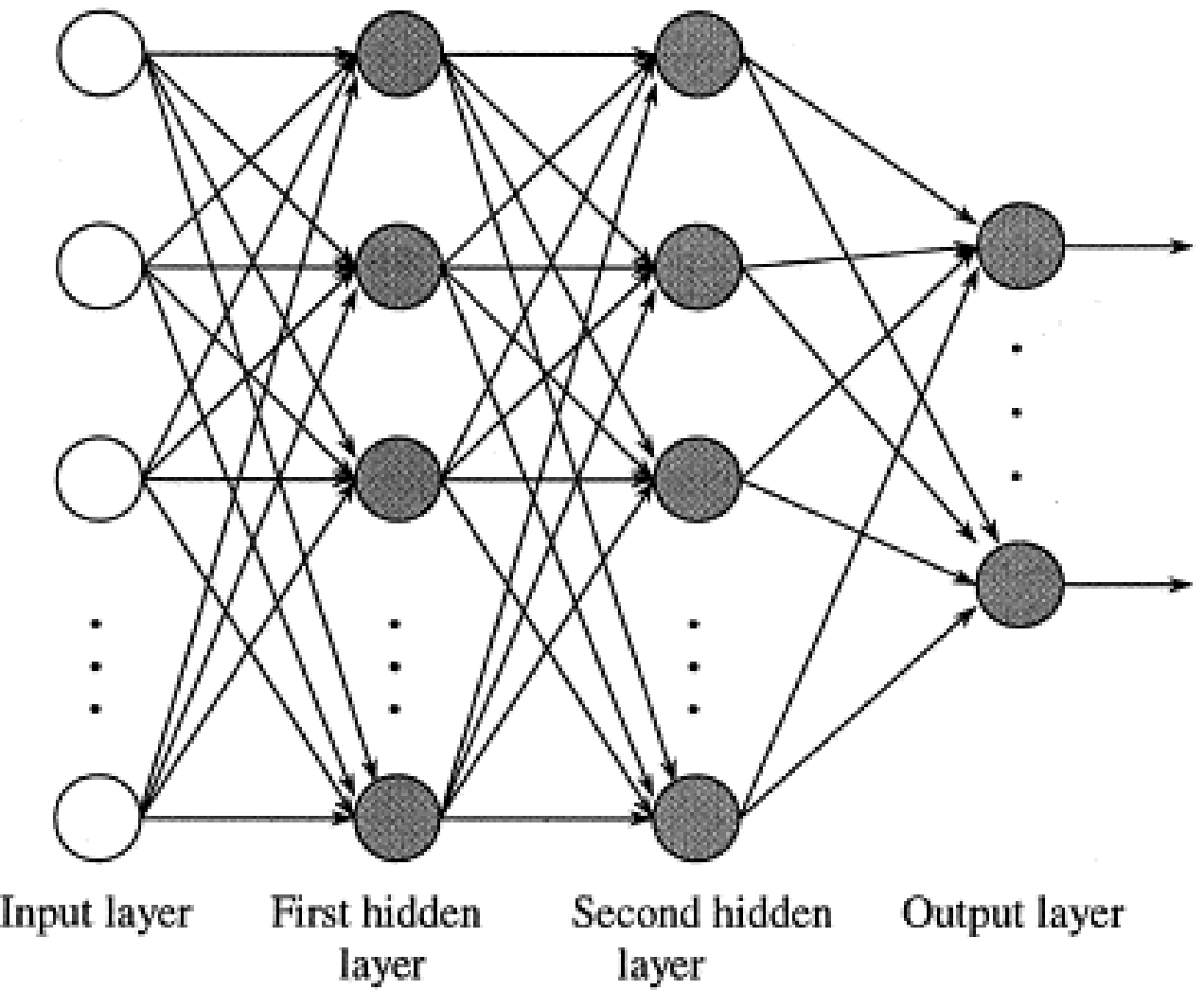}
\vskip 2truecm
\caption{
Conventional multilayer feedforward neural network (Multilayer 
Perceptron) having two intermediate layers of ``hidden neurons'' between 
input and output layers.  Darkened circles represent processing units 
analogous to neurons; lines oriented forward symbolize weighted 
connections between units analogous to interneuron synapses.  Information 
flows left to right as units in each layer simulate those in the next 
layer.} 
\end{figure}

\newpage

\begin{figure}
\includegraphics[width=.9\textwidth]{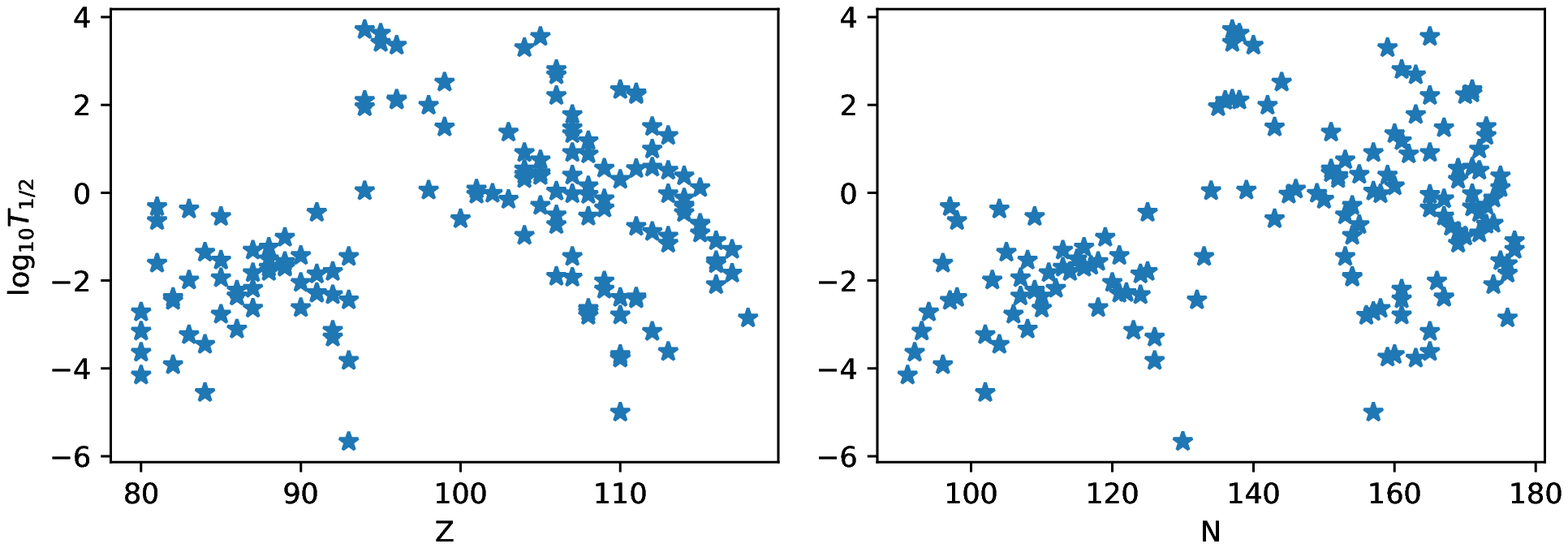}
\includegraphics[width=.9\textwidth]{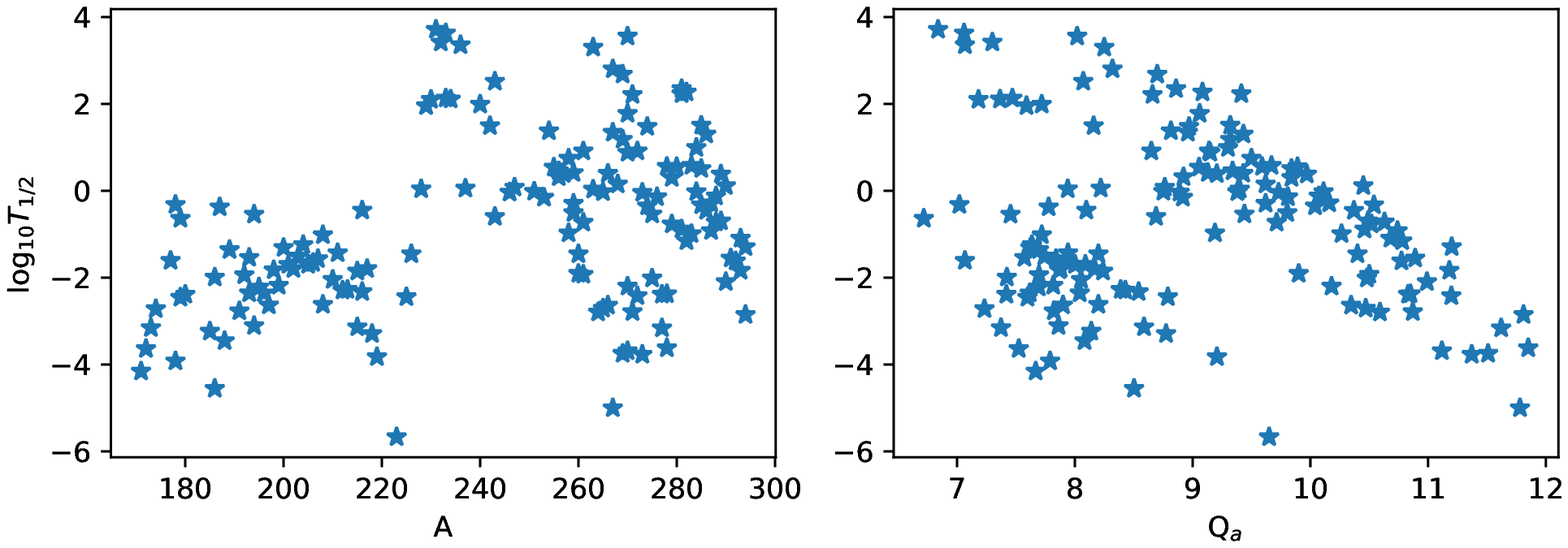}
\vskip 2truecm
\caption{
Visualizations of measured experimental alpha-decay half-lives 
(in $\log_{10}$ scale) for the 150 nuclei in the chosen database, 
plotted versus atomic number $Z$, neutron number $N$, mass 
number $A$, and energy release $Q_\alpha$.}
\end{figure}

\newpage

\begin{figure}
\vskip 2truecm
\includegraphics[width=.9\textwidth]{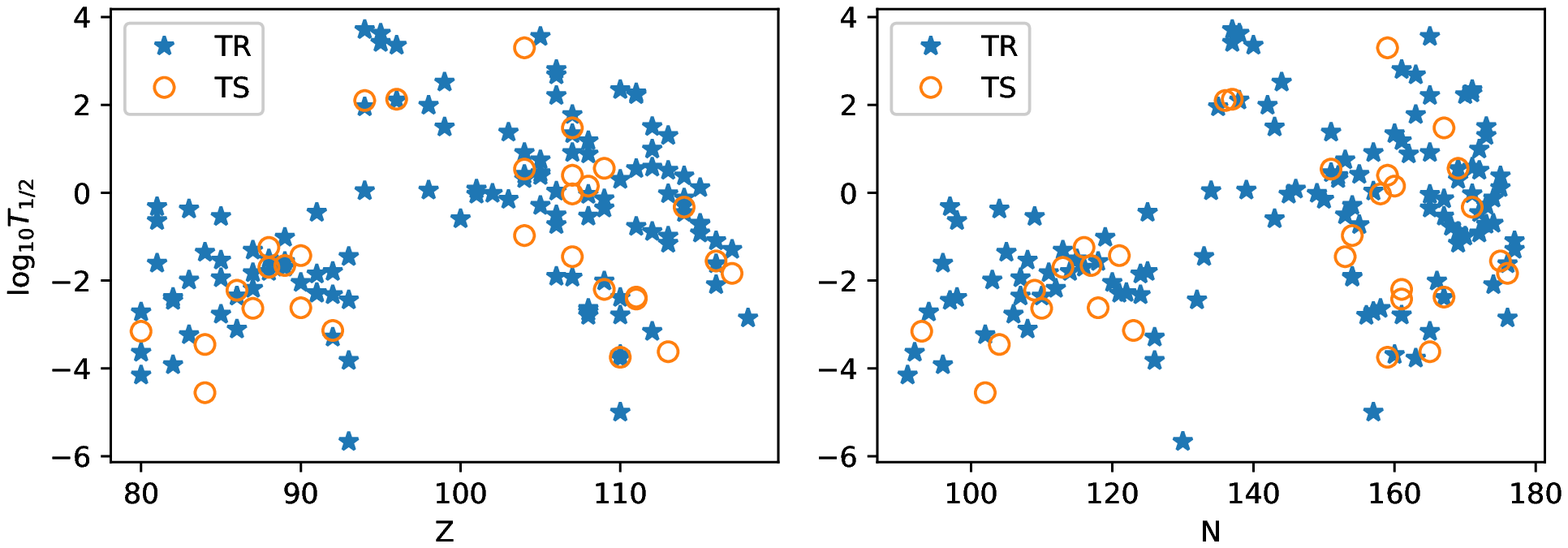}
\includegraphics[width=.9\textwidth]{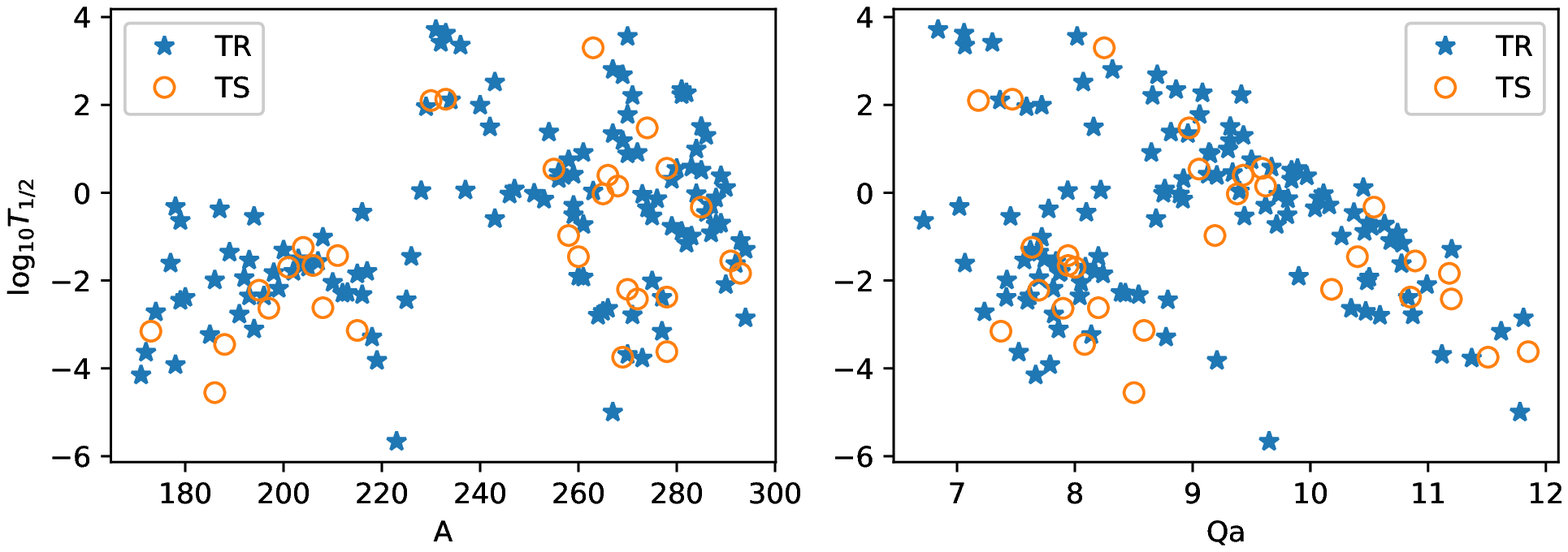}
\caption{
Visualizations of the training set (TR) and test set (TS) adopted
for {\bf Net1}, plotted with respect to atomic number $Z$, neutron 
number $N$, mass number $A$, and energy release $Q_\alpha$.}
\end{figure}

\newpage

\begin{figure}
\vskip 2truecm
\includegraphics[width=.9\textwidth]{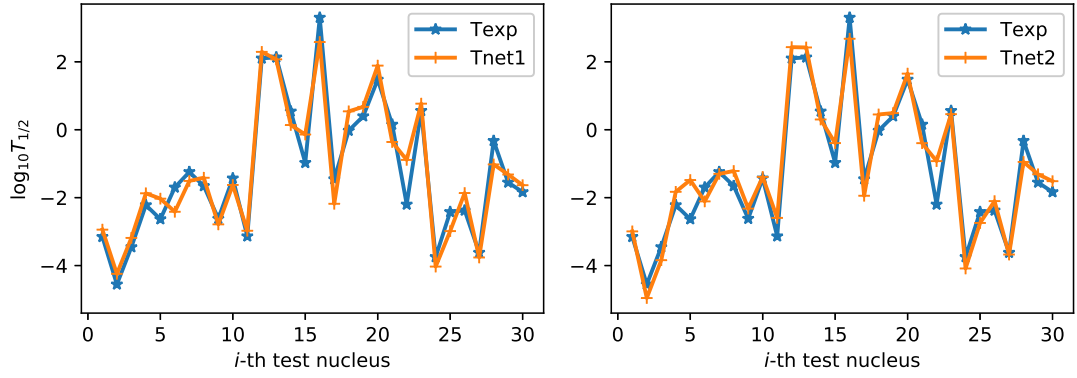}
\caption{
Schematic representation of the performance of {\bf Net1} and {\bf Net2}
in prediction of half-lives of test nuclides.}
\end{figure}

\newpage

\begin{figure}
\includegraphics[width=.9\textwidth]{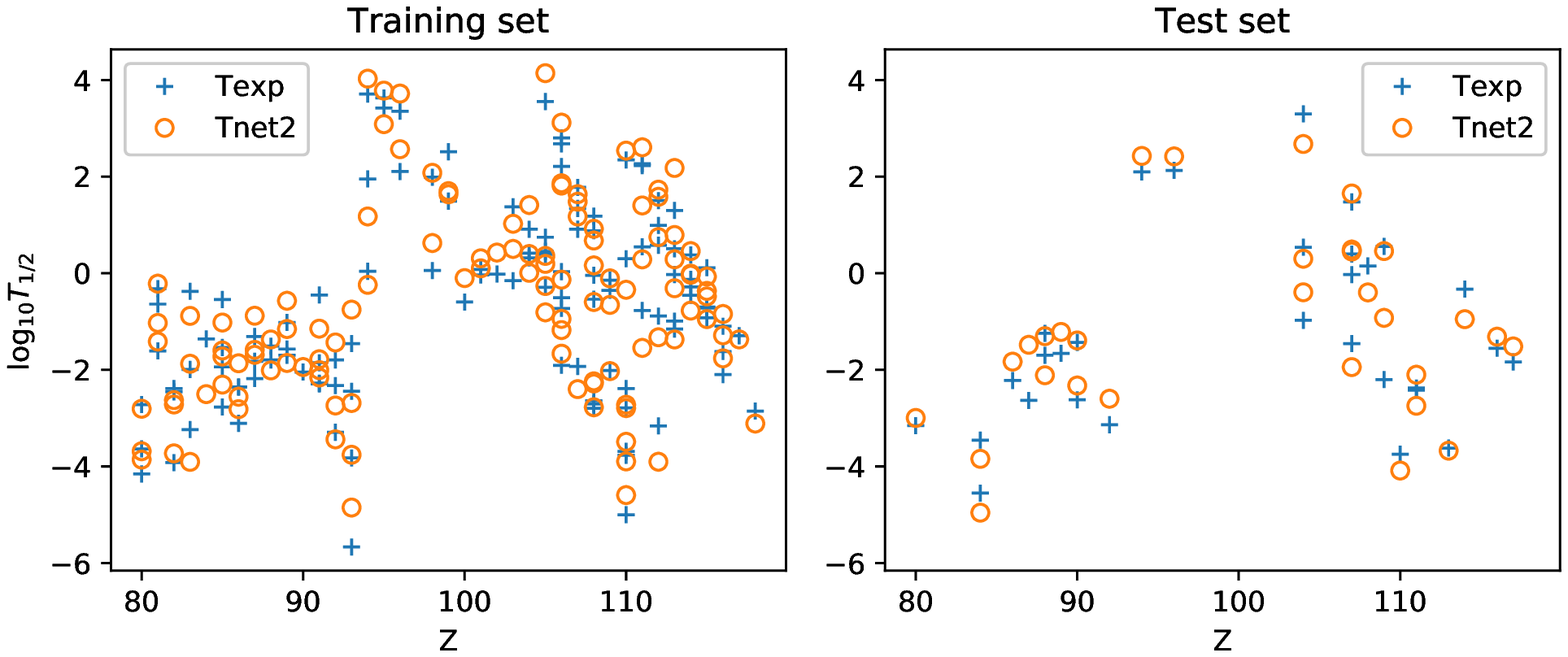}
\includegraphics[width=.9\textwidth]{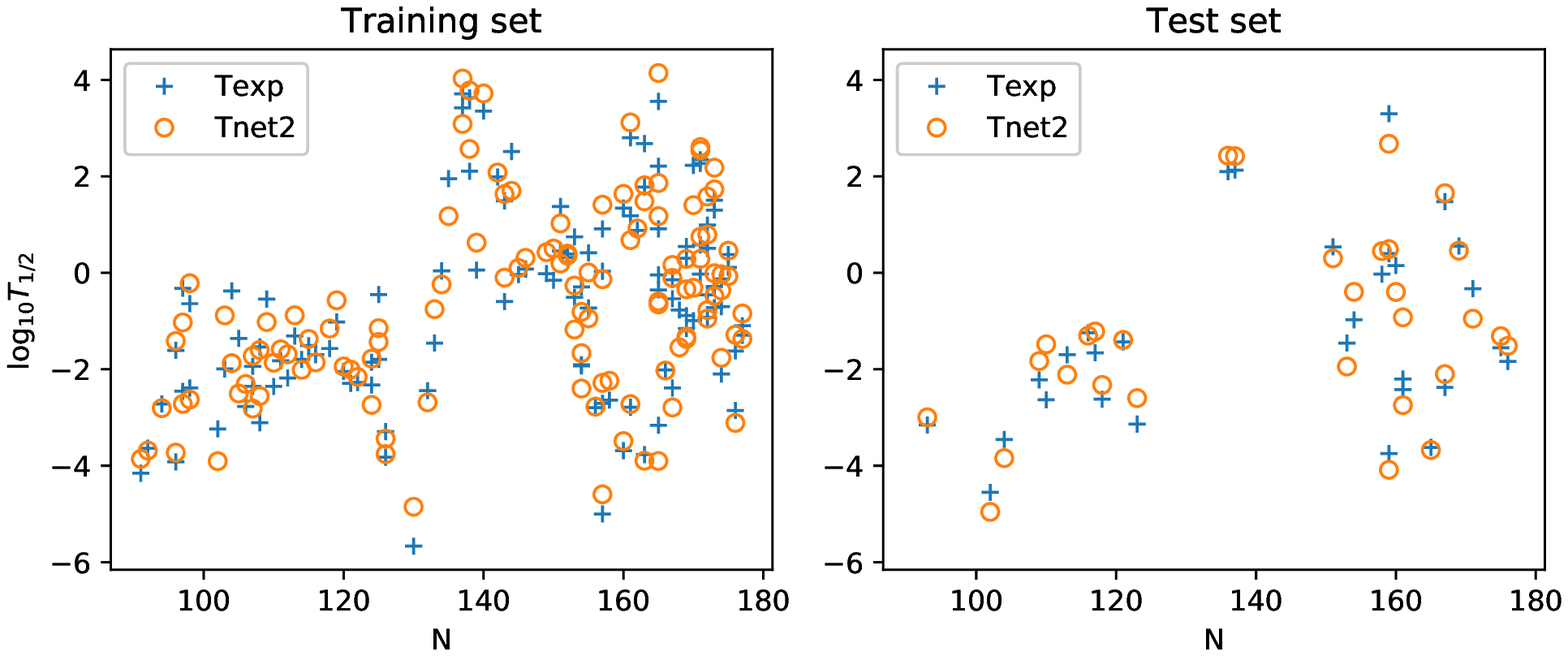}
\includegraphics[width=.9\textwidth]{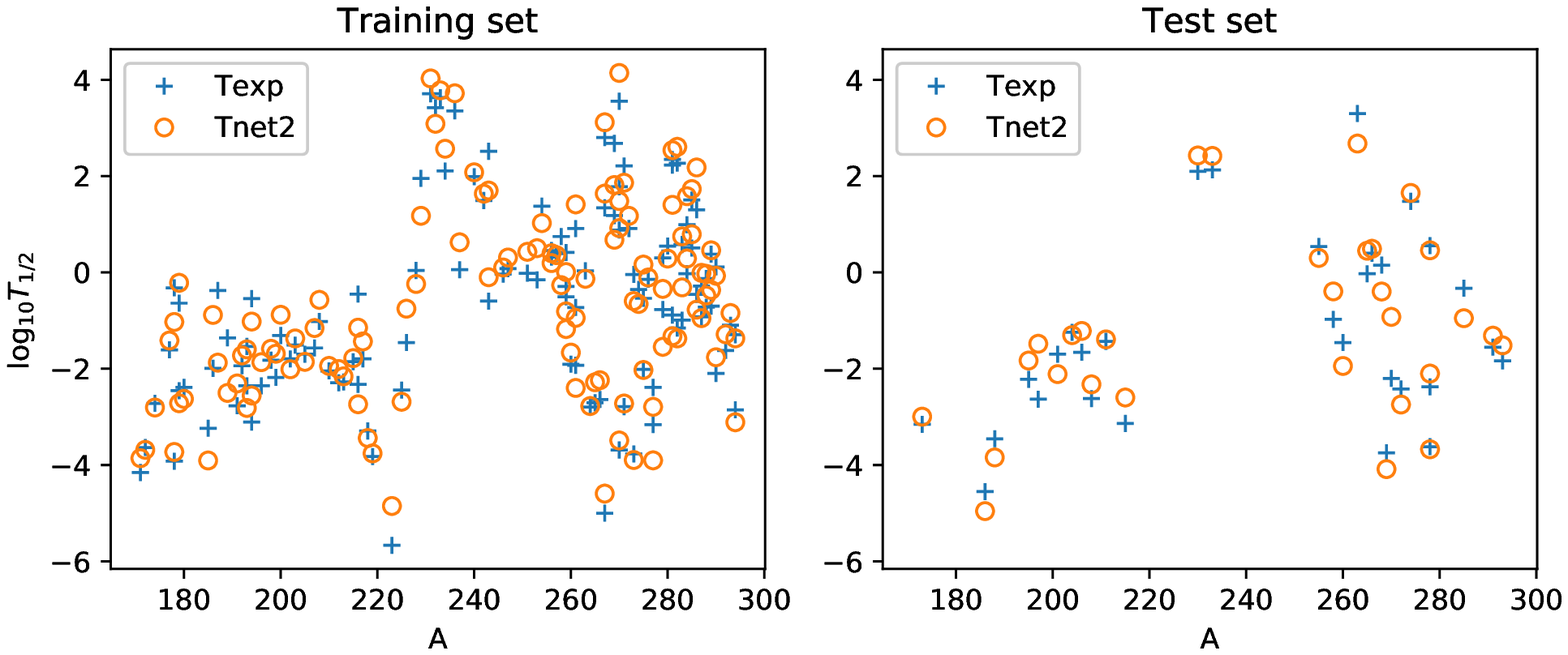}
\vskip 2truecm
\caption{
Ensembles of half-life fits for training nuclei and predictions for test
nuclei, generated by ${\bf Net2}$ and plotted versus atomic number $Z$, 
neutron number $N$, and mass number $A$.}
\end{figure}

\end{document}